\documentclass[a4paper,fleqn,usenatbib]{mnras}

\usepackage{newtxtext,newtxmath}

\usepackage[T1]{fontenc}
\usepackage{ae,aecompl}

\usepackage{graphicx}	
\usepackage{amsmath}	
\usepackage{amssymb}	

\title[Dust masses of $z>5$ galaxies]{Dust masses of $z>5$ galaxies
from SED fitting and ALMA upper limits}

\author[H. Hirashita, D. Burgarella, and R. J. Bouwens]{
Hiroyuki Hirashita,$^{1}$\thanks{E-mail: hirashita@asiaa.sinica.edu.tw}
Denis Burgarella$^{2}$ and Rychard J. Bouwens$^3$
\\
$^{1}$Institute of Astronomy and Astrophysics, Academia Sinica,
PO Box 23-141, Taipei 10617, Taiwan\\
$^{2}$Aix-Marseille Universit\'{e}, CNRS, LAM
(Laboratoire d'Astrophysique de Marseille) UMR 7326, 13388 Marseille, France\\
$^3$Leiden Observatory, Leiden University, NL-2300 RA Leiden, Netherlands
}

\date{Accepted XXX. Received YYY; in original form ZZZ}

\pubyear{2017}

\begin{document}
\label{firstpage}
\pagerange{\pageref{firstpage}--\pageref{lastpage}}
\maketitle

\begin{abstract}
We aim at constraining the dust mass in high-redshift ($z\gtrsim 5$)
galaxies using the upper limits obtained by ALMA in combination with
the rest-frame UV--optical spectral energy distributions (SEDs).
For SED fitting, because of degeneracy between dust extinction
and stellar age, we focus on two extremes: continuous star formation
(Model~A) and instantaneous star formation (Model~B). We apply
these models to Himiko (as a representative UV-bright object)
and a composite SED of $z>5$ Lyman break galaxies (LBGs).
For Himiko, Model~A {requires a significant dust extinction,
which leads to a high dust temperature $>70$ K for consistency with}
the ALMA upper limit.
{This high dust temperature puts
a strong upper limit on the total dust mass $M_\mathrm{d}\lesssim 2\times 10^6$ M$_{\sun}$,
and the dust mass produced per supernova (SN)
$m_\mathrm{d,SN}\lesssim 0.1$ M$_{\sun}$. Such a low $m_\mathrm{d,SN}$
suggests significant loss of dust by reverse shock destruction or outflow, and implies that
SNe are not the dominant source of dust at high $z$.}
Model B allows {$M_\mathrm{d}\sim 2\times 10^7$ M$_{\sun}$ and
$m_\mathrm{d,SN}\sim 0.3$ M$_{\sun}$.} We could distinguish between Models A and B
if we observe Himiko at {wavelength $<$ 1.2 mm} by ALMA.
For the LBG sample, we obtain $M_\mathrm{d}\lesssim 2\times 10^6$ M$_{\sun}$
for a typical LBG at $z>5$, but this only puts an upper limit for
$m_\mathrm{d,SN}$ as $\sim 2$ M$_{\sun}$.
This clarifies the importance
of observing UV-bright objects (like Himiko) to constrain
the dust production by SNe.
\end{abstract}




\section{Introduction}

Dust plays an important role in the evolution of galaxies and their interstellar
medium (ISM). Dust surfaces are
the main site for the formation of some molecular species, especially
H$_2$ \citep[e.g.][]{Gould:1963aa,Cazaux:2004aa}, inducing the formation
of molecular clouds, which host star formation
\citep[e.g.][]{Hirashita:2002aa,Yamasawa:2011aa}.
In the later stage of star formation, dust cooling induces fragmentation
\citep{Omukai:2005aa} and determines the typical stellar mass
\citep{Schneider:2006aa}.

Dust also modifies the appearance of galaxies by absorbing and
scattering stellar light
and reemitting it into far-infrared (FIR)\footnote{In this paper, we simply use
the term FIR for the wavelength range where the emission is dominated by
dust.} wavelengths. Therefore, dust dramatically
modifies the observed spectral energy distributions (SEDs) of galaxies
\citep[e.g.][]{Takeuchi:2005aa}. From a theoretical point of view,
consistent modelling of dust extinction (dust absorption and scattering) and
dust reemission is crucial to understand and constrain the dust
properties robustly \citep[e.g.][]{Calzetti:2001aa,Buat:2012aa}.
In other words, modelling only one of dust extinction
and dust emission is a highly degenerate problem as mentioned below.
Precisely speaking, we should refer to
the difference between the intrinsic stellar SED and the observed
SED as dust attenuation (not dust extinction), since
complex effects of radiation transfer in the galaxy also matters
\citep{Calzetti:2001aa,Inoue:2005aa}.
However, because there is no risk of confusion in this paper, we
simply use the term `extinction' without strictly distinguishing
between extinction and attenuation.

Dust extinction could be estimated to match the SED at
ultraviolet (UV) and optical wavelengths with a given stellar intrinsic SED.
However, it is generally difficult to separate the effect of
dust extinction and that of stellar age, since both effects
make the SED red. Moreover, the SED also depends on
the shape of extinction curve (i.e.\ the wavelength dependence of
dust extinction). Therefore, without any assumption on the
intrinsic stellar SED and extinction curve shape,
determining the age and extinction is a highly degenerate problem.

This degeneracy could be resolved at least partially
if we additionally use the FIR dust emission.
Because the stellar radiation energy absorbed by dust is
emitted in the FIR, the total FIR emission constrains the total
dust extinction. Indeed, this energy balance between dust
absorption and emission has been used to derive the total
dust extinction \citep{Buat:1996aa,Buat:1998aa,Takagi:1999aa}.
Therefore, this energy balance is the key to understand
the effect of dust extinction on the SED shape ranging from UV to FIR.

The link between UV extinction and FIR emission has been
investigated with the so-called IRX--$\beta$ relation, where the IRX
is the infrared excess (FIR-to-UV flux ratio) and $\beta$ is the UV
slope. This relation indicates that a large dust extinction
leads to a red UV SED and a high IRX \citep{Meurer:1999aa,Takeuchi:2012aa}.
Although the IRX--$\beta$ relation provides a powerful tool to
investigate the dust extinction and emission properties in galaxies,
we also observe a significantly different IRX--$\beta$ relation at
$z\gtrsim 5$ ($z$ is the redshift) from the one at low redshift
\citep{Capak:2015aa,Fudamoto:2017aa}.
Different extinction curves as well as multiple dust temperature structures
are possible reasons for the difference
\citep[][hereafter B16]{Mancini:2016aa,Ferrara:2017aa,Narayanan:2017aa,Bouwens:2016aa}.
Because the IRX--$\beta$ relation is not fully understood for high-$z$
galaxies, it is still worth investigating the UV--FIR SED directly.

Recently, it has become possible to investigate the dust production
and evolution in high-$z$ galaxies.\footnote{In this paper, we
refer to $z>5$ as high redshift.} To clarify the origin of dust
in the Universe, it would be desirable to observe the first-generation
galaxies, which is difficult at the current sensitivity of observational
facilities. The most sensitive dust search at high $z$ is possible by
the Atacama Large Millimetre/submillimetre Array (ALMA).
The highest-$z$ galaxies for which the dust emission is
detected by ALMA are located at $z>7$
\citep{Watson:2015aa,Willott:2015aa,Laporte:2017aa}.
The high sensitivity of ALMA enables us to constrain the dust
enrichment processes in those galaxies \citep{Mancini:2015aa,Wang:2017aa}.
However, dust continuum has not been detected for
most Lyman break galaxies (LBGs) at $z\gtrsim 6$
(B16; \citealt{Aravena:2016aa}).

\citet[][hereafter O13]{Ouchi:2013aa} observed
a Lyman $\alpha$ (Ly$\alpha$)-emitting gas blob `Himiko' at $z = 6.6$
using ALMA. O13 put an upper limit of
0.0521 mJy (3$\sigma$) at 1.2 mm for Himiko.
\citet[][hereafter H14]{Hirashita:2014aa} developed
a method to constrain the dust mass formed per SN based on O13's
result: They basically
divided the total dust mass by the total number of SNe estimated from
the UV luminosity, taking into account the fraction of dust destroyed
by SN shocks sweeping the ISM \citep[see also][]{Michalowski:2015aa}.
They obtained an upper limit of dust mass formed per SN for Himiko
as $\sim 0.15$--0.45 M$_{\sun}$ depending on
the assumed grain species. The obtained dust mass indicates that
a significant fraction of the dust once condensed in a SN is destroyed
in the shocked region before being injected into the ISM. This destruction
is referred to as reverse shock destruction.

The above analysis in H14 treated the UV and FIR SEDs separately.
As explained above, the SEDs in those two wavelength ranges are
tightly related through dust absorption and reemission.
It would be interesting to reexamine the above constraint on
the SN dust production by treating those two wavelength
ranges consistently with a single SED model. Therefore, in this paper,
we apply an SED model to the observed SED of Himiko
and reexamine the constraint on the dust mass. The advantage
of using such an SED argument is that we guarantee the energy
balance between absorption and reemission.

Although most LBGs at $z\gtrsim 6$ are not detected by ALMA, we expect
that we could obtain a stringent upper limit for the dust emission
of a typical LBG by stacking all the non-detections.
Therefore, another purpose of this paper is to constrain
the dust mass for high-$z$ LBGs.
Because the number of high-$z$ LBGs observed by ALMA is expected to increase,
the method developed here could also be applied to a larger sample in the future.
Because only a small number of LBGs are detected by ALMA,
we concentrate on the non-detections for the uniformity of the sample.
Detailed analysis of detected high-$z$ objects is given in our
separate paper (Burgarella et al., in preparation).

There are a large number of SED models based on stellar population synthesis
and dust attenuation treatment \citep[][for a review]{Conroy:2013aa}.
Some of them solve radiation transfer in a dusty ISM to obtain the SED
\citep[e.g.][]{Silva:1998aa,Takagi:2003aa,Bianchi:2008aa,Baes:2011aa,Popescu:2011aa,De-Looze:2014aa,Yajima:2014aa}.
Although radiation transfer modelling enables us to take realistic spatial
distributions of dust and stars into account, it generally has a high computational
cost. Moreover, little is known about the geometry of dust and stellar
distributions for high-$z$ galaxies, which means that apparent morphologies cannot
be used to constrain the model. Given the situation, a simple SED model
that is computationally less expensive but still considers the energy
balance between dust extinction and emission is useful for high-$z$ galaxies.
In this case, instead of solving radiation transfer, we treat the galaxy
as a single-zone object, but
we are able to run a lot of cases for different dust extinctions, dust
properties (especially, extinction curves), and stellar population ages.
There are some SED models suitable for such a purpose
\citep{Burgarella:2005aa,da-Cunha:2008aa}. The basic idea of these
models is that the stellar light is synthesized based on the star formation
history, attenuated according to the assumed dust extinction curve, and
reemitted in the FIR. Among them, we adopt \textsc{cigale}
\citep{Noll:2009aa}, but the results in
this paper will not be changed even if we adopt other SED models.

This paper is organized as follows.
In Section \ref{sec:model}, we explain the observational data
and the SED model. In Section \ref{sec:result}, we show the results
of the SED fitting and dust mass estimates.
In Section \ref{sec:SNdust}, we constrain the dust production rate
by SNe based on the results.
In Section \ref{sec:discussion}, we discuss the limitation of our
method and the implication of our results for dust enrichment at
high $z$. In Section \ref{sec:conclusion}, we give the conclusion of
this paper. We use
$(h,\,\Omega_\mathrm{m},\,\Omega_\Lambda )=(0.7,\, 0.3,\, 0.7)$
for the cosmological parameters.

\section{SED fitting}\label{sec:model}

\subsection{Data}\label{subsec:data}

For the direct comparison with H14, we adopt Himiko
(a large Ly$\alpha$ emitting galaxy) to constrain the dust mass.
Since Himiko is one of the brightest galaxies in the rest UV at $z>6$
but is not detected by ALMA, it has potentially very low FIR-to-UV luminosity ratio.
This leads to a stringent limit for the dust production in the early
epoch of galaxy evolution.
We adopt the rest-frame UV--optical SED data in
\citet{Ouchi:2013aa} for Himiko.

Other than Himiko, there have been a lot of high-$z$ galaxies, mainly LBGs,
observed by ALMA. One of the largest
samples can be found in B16. None of the sample LBGs at $z>5$ in B16
was detected by ALMA.
Although each LBG gives only a weak constraint on the dust production
compared with Himiko, stacking the large sample could enable us to
obtain a strong constraint on the dust mass. In fact,
a small number of LBGs are detected by ALMA at $z>5$
\citep{Capak:2015aa,Watson:2015aa,Laporte:2017aa}. Since these galaxies
need SED fitting and detailed analysis one by one, we treat them in
our future paper (Burgarella et al., in preparation). We emphasize that
a major part of LBGs are not detected by ALMA and that they can
be analyzed uniformly with our method developed in this paper.
The methodology established here could be applied to any future larger sample.
For the first step, we produce a composite (or stacked) SED for the B16 sample.

We select galaxies at $z\geq 5.0$ in B16 (78 objects in total).
The redshift of the sample extends up to $z=9.8$, and most
of the sample are located at $z<7$ with 13 exceeding $z=7$.
First of all, we need to make the redshifts uniform. In principle we can
choose any redshift; however, for this paper, it is convenient to choose
Himiko's redshift ($z=6.6$) since we can utilize the same SED models.
For {an} LBG at original redshift $z_\mathrm{orig}$,
we move it virtually to $z=6.6$ by shifting the wavelength by
$\lambda (1+z)/(1+z_\mathrm{orig})$ and multiplying the flux with
$[(1+z)/(1+z_\mathrm{orig})][d_\mathrm{L}(z_\mathrm{orig})/d_\mathrm{L}(z)]^2$,
where $d_\mathrm{L}(z)$ is
the luminosity distance at redshift $z$ \citep{Carroll:1992aa}.

\begin{figure}
 \includegraphics[width=0.95\columnwidth]{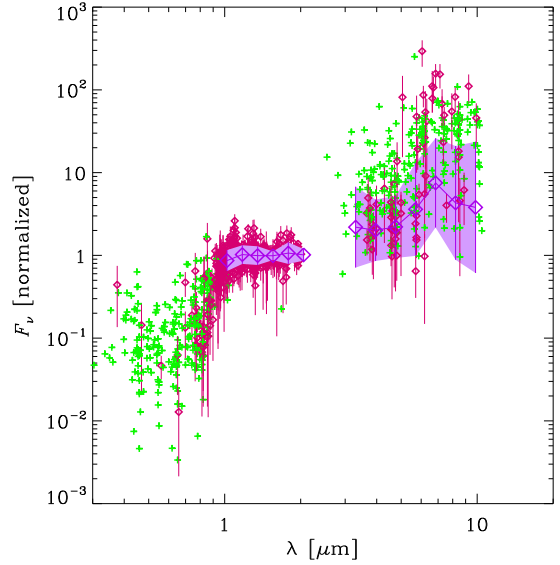}
 \caption{SEDs of the B16 LBG sample shifted to $z=6.6$ and
 normalized to the flux at rest 0.2 $\micron$ (i.e.\ 1.52 $\micron$
 in this figure). The wavelength in this figure (and also in the following
 figures) is shown in the observer's frame.
 The small squares with error bars show detected
 data points while the crosses show upper limits.
 The large squares show the composite SED with the shaded regions
 corresponding to the dispersion for the data at $\lambda <3~\micron$
 and to the upper and lower ranges for the data at $\lambda >3~\micron$
 (see the text for details). The normalization of the composite SED at 1.52 $\micron$ is
 determined as 9.3 nJy.}
 \label{fig:LBGSED}
\end{figure}

Next, since the flux level is diverse among the sample LBGs,
{we need to normalize the SED at a certain wavelength,
in order to} extract the information on the SED shape.
For this purpose, we normalize the flux to
the value at rest 0.2 $\micron$ (i.e.\ 1.52 $\micron$ after shifting
the SED).
The flux at rest 0.2 $\micron$ is estimated from the flux at the
two nearest wavelengths by
interpolation or extrapolation and divide the fluxes at all the
sampled wavelength by the 0.2-$\micron$ flux.
The shifted and normalized fluxes for all the sample are plotted in
Fig.\ \ref{fig:LBGSED}.

Based on the above normalized SEDs of the sample, we make a
composite SED. As we observe in Fig.\ \ref{fig:LBGSED}, the
SEDs are roughly divided into the three parts: at
$\lambda <0.92~\micron$, most of the points are upper limits
because the radiation is absorbed by hydrogen atoms in
the interstellar or intergalactic medium on the line of sight.
We do not use the data in this wavelength range for the fitting below.
At $0.92<\lambda <3~\micron$, most of the points are detected
data; thus, we neglect the data without detection in this wavelength range.
We divide the data into 6 bins with a logarithmically equal width
(0.95--1.09, 1.09--1.26, 1.26--1.44, 1.44--1.66, 1.66--1.91, and
1.91--2.20 $\micron$).
We take the average of
the logarithmic values of the normalized data points
in each bin to obtain the composite SED.
We also estimate the logarithmic dispersion as shown in Fig.\ \ref{fig:LBGSED}.
At $\lambda >3~\micron$, a large fraction of the data points are
not detected. In this wavelength range, thus, we would overestimate
the averaged flux if we neglect all the points without detection.
{To avoid such an overestimate, we derive the probable range
of the composite SED at $\lambda >3~\micron$ by estimating upper and lower
bounds in the following way. First, we
set 7 bins with a logarithmically equal width (3.00--3.60,
3.60--4.32, 4.32--5.18, 5.18--6.22, 6.22--7.46, 7.46--8.96, and 8.96--10.7 $\micron$).
We average the logarithmic upper and lower bounds
of the sample to obtain the upper and lower limits, respectively, in each wavelength bin.
For detected data points, we simply use the
observed flux for both upper and lower bounds.
For non-detections, we use the upper limit given by B16, while
we give a lower limit by adopting the value estimated from
the bluest SED as explained below.
The bluest SED under a fixed $\lambda =1.52~\micron$ (rest 0.2 $\micron$)
flux gives
the lowest possible flux at $\lambda >3~\micron$, and can be estimated by
extrapolating the rest 0.2 $\micron$ flux (i.e.\ unity after the normalization)
with a power-law $f_\nu\propto\lambda^{\beta_\mathrm{UV}+2}$,
where $\beta_\mathrm{UV}$ is the spectral slope at rest UV wavelengths
\citep{Calzetti:1994aa}. If we adopt the smallest possible value of
$\beta_\mathrm{UV}$, we obtain a lower limit of the flux.
The smallest value of $\beta_\mathrm{UV}$ is determined by the
intrinsic stellar UV SED; thus, we adopt $\beta_\mathrm{UV}=-2.5$
\citep{Bouwens:2014aa}.
For the fitting, we allow for the full range between the upper and lower
limits and} adopt the logarithmic average of these two limits as the representative
value in each wavelength bin.
Finally, the normalized flux is multiplied by the logarithmically averaged
rest 0.2 $\micron$ flux (9.3 nJy) to obtain the absolute level of the flux.

The upper limits of the 1.2 mm flux given by B16 are utilized to obtain the
upper limit of the millimetre (mm) flux for the composite (or stacked) SED.
First, we obtain a stacked upper limit by
$\bar{f}_\mathrm{mm}=\sqrt{\sum_i\sigma_i^2}/N$, where
$\sigma_i$ is the 1$\sigma$ noise level for the $i$th galaxy.
Next, we calculate the mean redshift $\bar{z}=6.1$ for the B16 sample.
The wavelength (1.24~mm) and the upper limit flux
$\bar{f}_\mathrm{mm}$ are shifted in the same way as above
to obtain the corresponding values
at $z=6.6$. Consequently, we obtain a 3$\sigma$ upper limit value of
6.7 $\mu$Jy at 1.31~mm.

\subsection{SED code -- \textsc{cigale}}\label{subsec:cigale}

We use \textsc{cigale} (Code Investigating GALaxy Emission)
\citep{Noll:2009aa} to produce the
UV--submm SED of a galaxy. It takes into account the energy balance
between the stellar light extinguished by dust and its re-emission in the
FIR.  The stellar population synthesis is based on
\citet{Bruzual:2003aa} and \citet{Maraston:2005aa}.
There are some freedoms in the parameter setting, which we set
as described below.

We adopt a Chabrier initial mass function (IMF) \citep{Chabrier:2003aa},
although applying a Salpeter IMF \citep{Salpeter:1955aa} instead
does not change our results below significantly.
We also include emission lines,
since they are known to contribute to the fluxes in some bands for
star-forming galaxies.
We use the following functional form for the star formation rate (SFR):
\begin{align}
\mathrm{SFR}(t)=C(t/\tau_\mathrm{SF})\exp(-t/\tau_\mathrm{SF}) ,
\end{align}
where $C$ (proportional to the total stellar mass, $M_*$)
is the normalization constant adjusted in the fitting,
$t$ is the age, and $\tau_\mathrm{SF}$ is the star formation
time-scale.
This functional form can mimic a continuous (or constant) SFR if we adopt
a much longer $\tau_\mathrm{SF}$ than the age
($\tau_\mathrm{SF}\gg t$) while it represents a burst SFR if
we take $\tau_\mathrm{SF}\ll t$.

For the dust SED, we adopt the \textsc{casey2012} module, which
is based on \citet{Casey:2012aa}. In this SED model, the FIR emission
is practically the so-called modified black body radiation which is
described by a functional form of $\nu^\beta B_\nu (T_\mathrm{d})$,
where $\nu$ is the frequency, $T_\mathrm{d}$ is the dust temperature,
and $B_\nu (T_\mathrm{d})$ is the Planck function.
We are not interested in the power-law like mid-infrared emission
adopted in the \textsc{casey2012} module
in this paper. The advantage of this model is that we are able to give
the dust temperature freely. We fix $\beta =1.6$, but 
this choice does not affect the results below significantly as long
as we adopt $\beta =1$--2.

For the extinction law, we adopt the power-law form, since the
detailed functional form is not important (and cannot be constrained)
in this work. In this model, the extinction at wavelength $\lambda$ is
described by a given power-law index $\delta$ as
\begin{align}
A_\lambda =A_V\left(\frac{\lambda}{0.55~\micron}\right)^\delta ,
\end{align}
where $A_V$ is the extinction in the $V$ band. We examine the
following three cases for $\delta$: $\delta\simeq -0.4$,
$-0.7$, and $-1.1$, which roughly approximate
a flat extinction curve as observed in a high-redshift
quasar by \citet{Maiolino:2004aa} and \citet{Gallerani:2010aa},
an attenuation curve representative of nearby starburst galaxies
\citep{Calzetti:1994aa}, and the Small Magellanic Cloud (SMC) extinction
curve \citep{Pei:1992aa}, respectively.
Note that the flat extinction curves in high-$z$ quasars are
also consistent with theoretically expected dust properties
for SN dust production \citep{Maiolino:2004aa,Hirashita:2005ab,Asano:2014aa}
or for strong grain growth by coagulation \citep{Nozawa:2015aa}.

We also apply different extinctions ($A_V$) for the young ($<10$ Myr) and
old ($>10$ Myr) stellar populations following \citet{Charlot:2000aa}.
We denote the extinction of the young population in the $V$ band as
$A_V$. We introduce a parameter $\eta\leq 1$ that expresses the
extinction of the old population relative to that of the young population
(i.e.\  the extinction of the old stellar population is $\eta A_V$).
According to \citet{Calzetti:2001aa}, $\eta\simeq 0.44$ for
nearby star-forming galaxies. We also
examine other values such as
$\eta =0$ (no extinction for the stellar population with age $>10$ Myr)
and $\eta =0.9$ (almost no difference between the extinctions of
stellar populations with different ages).

For each parameter set $(A_V,\, T_\mathrm{d},\, \delta ,\,\eta )$,
we obtain {an} SED, and scale the
total stellar mass ($M_*$) to minimize $\chi^2$. The $\chi^2$ is estimated
using the logarithmic fluxes and errors (for the stacked LBG SED,
we use the half width of the
shaded range in Fig.\ \ref{fig:LBGSED} for $\sigma$ at each wavelength bin).
Although we do not use the ALMA upper limit for the fitting directly, we
only accept the case in which the
model flux at the ALMA band is below the 3$\sigma$ upper limit.
For Himiko, we choose the parameter sets of satisfactory fit
based on a criterion of reduced $\chi^2<3$. We have confirmed that
adopting more relaxed criterion
as $\chi^2<5$ does not change the results below (in other words,
the range of the acceptable parameter values does not become significantly wider).
For LBGs, we also minimize $\chi^2$ but only used the stacked data
at $0.92~\micron<\lambda <3~\micron$, and exclude the SEDs which
are not within the dispersion shown in Fig.\ \ref{fig:LBGSED} in that wavelength range.

\subsection{Extracting dust-related parameters}\label{subsec:strategy}

First, we performed fitting to the observed SED by
freely varying relevant parameters in \textsc{cigale}.
Overall, most of the parameters
are not constrained mainly because of the well-known degeneracy
between dust extinction and age, both of which contribute to the
`reddening' of the UV--optical SED. Only the age has a significant
range of $440\pm 250$~Myr, which is only weakly constrained though.
Similar stellar ages are also obtained by \citet{Ouchi:2013aa}.
The stellar metallicity is not constrained; thus, we fix it to
0.004 ($\sim$1/5~Z$_{\sun}$) throughout this paper.

Considering the age--extinction degeneracy, we choose to focus on
the two extreme (but still reasonable) cases:
(A) blue stellar continuum with high extinction, and
(B) red stellar continuum with low extinction.
These two cases are differentiated by the star formation
time-scale $\tau_\mathrm{SF}$. For (A), we adopt
$\tau_\mathrm{SF}=2000$~Myr and $t=400$~Myr.
This case represents a gradually rising
($\mathrm{SFR}\propto t/\tau_\mathrm{SF}$) star formation history
up to the age $t=400$~Myr
(consistently with the above age constraint; note that the mean age of the
stellar population is roughly 200 Myr in this case). This case is referred to as
the continuous SFR.
For (B), we adopt $\tau_\mathrm{SF}=20$ Myr, and $t=200$ Myr.
Since $\tau_\mathrm{SF}\ll t$, the mean stellar age is $\sim t=200$ Myr,
which is equal to the mean age of (A). This case is referred to as the
burst SFR. The value of $\tau_\mathrm{SF}$ is
chosen for the following reason. If $\tau_\mathrm{SF}\gtrsim 30$ Myr,
the contribution from the young ($<10$ Myr) population is not negligible.
In this case, the situation is similar to Model A and we need to include
an appreciable amount of extinction. In contrast,
$\tau\lesssim 10$ Myr is rejected since  the intrinsic stellar SED is significantly redder than the
observed SED of Himiko. Thus, we adopt $\tau =20$ Myr to represent the case of red intrinsic stellar
SED.

After fixing the star formation history, the remaining parameters
that dominate the SED are those related to the extinction and emission
of dust. Thus, the extinction $A_V$, which determines the energy
emitted in the FIR, and $T_\mathrm{d}$, which regulates the peak
wavelength of the FIR SED,  are the most important parameters.
As we will see later, further details of extinction also affect the results:
in particular, $\delta$ and $\eta$ regulate the SED colour
in the UV--optical (smaller $\delta$ makes the stellar SED redder under
a fixed $A_V$).
As shown later, since $\delta$ and $\eta$ show different effects on
the stellar and dust SEDs, we vary both {of} those parameters.

In summary, we
vary $A_V$, $T_\mathrm{d}$, $\delta$, and $\eta$ in searching for
a fitting solution. For $\delta$ and $\eta$, unless otherwise stated,
we consider the following
representative cases: $\delta =-0.4$, $-0.7$, and $-1.1$;
$\eta =0$, 0.44, and 0.9 (see Section \ref{subsec:cigale}).
We move $A_V$ and $T_\mathrm{d}$ (quasi) continuously.
We basically apply the same procedure for both Himiko and
the B16 LBG sample.

\subsection{Constraint on the dust mass}\label{subsec:Mdust}

The total FIR luminosity (= total stellar radiation energy extinguished by dust)
and the dust temperature can be translated into the total dust mass.
The flux density at frequency $\nu$ in the observer's frame can be written
as (H14)
\begin{align}
f_\nu =
\frac{(1+z)\kappa_{(1+z)\nu}M_\mathrm{d}B_\mathrm{(1+z)\nu}(T_\mathrm{d})}{d_\mathrm{L}^2},
\label{eq:Mdust}
\end{align}
where $\kappa_\nu$ is the dust mass absorption coefficient at
$\nu$, and $M_\mathrm{d}$ is the dust mass. H14 give
the dust mass absorption coefficient
at $\lambda =158~\micron$ ($\kappa_{158}$), which
{corresponds to the ALMA-observed wavelength at the redshift of Himiko}
($z=6.6$; 1.2~mm in the observer's frame).
The mean wavelength of the B16 sample is 1.31~mm. The difference
in the wavelength is corrected for $\kappa_\nu$ by assuming a dependence of
$\kappa_\nu\propto\nu^\beta$ with $\beta =1.6$ (i.e.\ consistent dependence
with the SED fitting). Because this correction is small,
the detailed wavelength dependence of
$\kappa_\nu$ does not influence our results.
We adopt the same dust species as in H14, and list the adopted values
of $\kappa_\nu$ at 158 $\micron$ for each
dust species in Table~\ref{tab:kappa}.
Among the various dust species, silicate and graphite are used to
model the extinction curves in nearby galaxies \citep{Draine:1984aa,Pei:1992aa,Hou:2016aa}.
Since SNe may contribute to the quick dust enrichment in $z>5$ galaxies
\citep{Todini:2001aa,Nozawa:2003aa,Maiolino:2004aa},
we also use the mass absorption coefficient for the dust grains formed in SNe.
We adopt the theoretically calculated mass absorption coefficient
for dust condensed in SNe (SN$_\mathrm{con}$), which was
obtained by \citet{Hirashita:2005ab} using the dust species and
grain size distribution in \citet{Nozawa:2003aa}. We also apply the dust properties after
the so-called reverse shock destruction within the SN remnant
\citep{Nozawa:2007aa,Hirashita:2008aa} (SN$_\mathrm{dest}$).
In addition, we examine amorphous carbon (AC), which was used to
model the SED of SN 1987A by \citet{Matsuura:2011aa}.

\begin{table}
\centering
\begin{minipage}{70mm}
\caption{Dust properties.}
\label{tab:kappa}
    \begin{tabular}{lccc}
     \hline
     Species & $\kappa_{158}\,^\mathrm{a}$ & $\kappa_\mathrm{sil}^{-1}\,^\mathrm{b}$ &
     Ref.\,$^\mathrm{c}$\\
      & (cm$^2$ g$^{-1}$) &\\
     \hline
     Graphite & 20.9 & 0.63 & 1, 2\\
     Silicate & 13.2 & 1 & 1, 2\\
     SN$_\mathrm{con}\,^\mathrm{d}$ & 5.57 & 2.4 & 3\\
     SN$_\mathrm{dest}\,^\mathrm{e}$ & 8.94 & 1.5 & 4\\
     AC$^\mathrm{f}$ & 28.4 & 0.46 & 5, 6\\
     \hline
    \end{tabular}
    
    \medskip

$^\mathrm{a}$Mass absorption coefficient at 158 $\micron$.\\
$^\mathrm{b}$Inverse of $\kappa_{158}$ normalized to the silicate value.
The dust mass obtained in this paper is basically for silicate; thus, if we multiply
the dust mass with $\kappa_\mathrm{sil}^{-1}$, we obtain the dust mass for
other dust species.\\
$^\mathrm{c}$ References: 1) \citet{Draine:1984aa}; 2) \citet{Dayal:2010aa};
3) \citet{Hirashita:2005ab}; 4) \citet{Hirashita:2008aa}; 5) \citet{Zubko:1996aa};
6) \citet{Zubko:2004aa}.\\
$^\mathrm{d}$Dust condensed in SNe before {reverse} shock destruction.\\
$^\mathrm{e}$Dust ejected from SNe after {reverse} shock destruction.\\
$^\mathrm{f}$Amorphous carbon.
\end{minipage}
\end{table}

\section{Results}\label{sec:result}

\subsection{Himiko: Model A}\label{subsec:ModelA}

For Model A ($\tau_\mathrm{SF}=2000$ Myr and $t=400$ Myr),
because of the blue intrinsic stellar SED, dust extinction is strongly
required.
However, a large extinction also indicates a high dust FIR luminosity;
in particular, Himiko has a very bright UV luminosity, which would lead to a
high FIR luminosity even for a small amount of extinction.
Because of the stringent upper limit at 1.2 mm, the smallest $\eta$ and
$\delta$ (i.e.\ $\eta =0$ and $\delta =-1.1$) give the most relaxed condition
for the extinction; that is, this choice of $\eta$ and $\delta$
minimizes the FIR emission by restricting the extinction only to the
youngest ($<10$ Myr) population
and by reddening the UV SED most efficiently with the steepest
extinction curve.

\begin{figure}
 \includegraphics[width=\columnwidth]{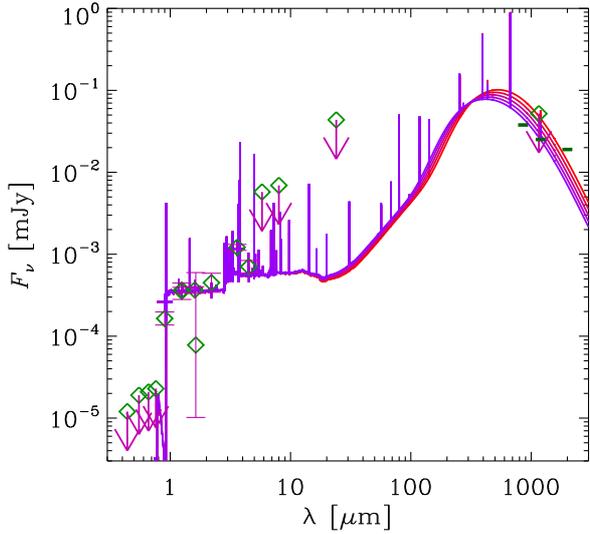}
 \caption{Examples of satisfactory fitting to the SED of Himiko in Model~A.
 We adopt $A_V=0.5$ {mag}, $\eta =0$ and $\delta =-1.1$. The SEDs are shown
 for $T_\mathrm{d}=70$, 75, 80, 85, and 90 K from
 the upper to lower lines at 1.2 mm. The data points (diamonds) are taken from
 \citet{Ouchi:2013aa}. The data points with error bars show detected points, while
 those with arrows indicate upper limits. The short horizontal lines at 0.85, 1.2, and 2 mm
 show the 3 $\sigma$ detection limits for 5-hour on-source integration
 by the full ALMA (50 12-m antennas).}
 \label{fig:Himiko_ModelA}
\end{figure}

With $\eta =0$ and $\delta =-1.1$, we find that the satisfactory fit
solutions have the following properties. Only dust temperatures higher than $\sim 70$ K
are consistent with the ALMA upper limit.
This is because, with a fixed FIR dust luminosity, the dust SED peak shifts
to a shorter wavelength and the 1.2 mm flux becomes lower
for a higher dust temperature.
We also find a satisfactory fit for any value of $A_V\geq 0.5$ {mag}
since the stellar SED
is consistent with the sum of the populations with age $>$ 10 Myr
(recall that, with $\eta =0$, we applied dust extinction only to stellar populations with
age $<$ 10 Myr).
This means that it is important to extinguish the radiation from
the youngest ($<$ 10 Myr) population. Thus, as examples of satisfactory fits,
we show the cases of $A_V = 0.5$ {mag}, $\eta =0$, and $\delta =-1.1$ with various
dust temperatures in Fig.\ \ref{fig:Himiko_ModelA}.

In Fig.\ \ref{fig:Himiko_ModelA}, we also show the ALMA sensitivities
expected for 5-hour on-source integration with the full ALMA
(50 12-m antennas) at wavelengths 0.85,
1.2 and 2 mm (frequencies 350, 250, and
150 GHz).\footnote{https://almascience.nao.ac.jp/proposing/sensitivity-calculator}
We observe that, because the required dust temperature
is high, the SED falls steeply toward long wavelengths.
As a consequence, a shorter wavelength band tends to
detect Himiko more easily. It is expected that
Himiko is detected at both 0.85 mm and 1.2 mm with the future
full ALMA sensitivity, if Model A is appropriate for Himiko.
{Since the 0.85 mm band is near to the SED peak,}
detection at two wavelengths {including 0.85 mm} enables
us to estimate the dust temperature and the total FIR luminosity
{under a given emissivity index ($\beta$)}.
The total FIR luminosity constrains the total amount of dust extinction
(i.e.\ $A_V$).
The non-detection at
$\sim 2$ mm would also confirm a high dust temperature.

\begin{figure}
 \includegraphics[width=\columnwidth]{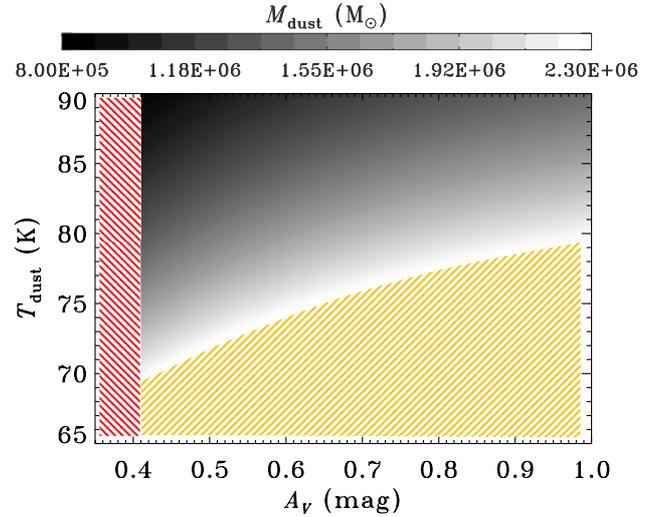}
 \caption{Dust mass derived from parameter set ($A_V$, $T_\mathrm{d}$)
 for Himiko in Model A.
 {The dust mass corresponding to ($A_V$, $T_\mathrm{d}$) is shown by
 the grey scale,}
 and the level of dust mass is shown
 in the bar on the top.
 {The dust mass is only shown in the area where the SED fitting is
 successful. The yellow shaded area is the region where the mm flux exceeds the ALMA
 upper limit, while the red shaded area is the region where we do not obtain
 a satisfactory fit to the rest UV--optical data} (i.e.\ reduced $\chi^2\geq 3$).}
 \label{fig:Mdust_ModelA}
\end{figure}

For a given set of ($A_V$, $T_\mathrm{d}$), we obtain the
dust mass using the conversion from the predicted mm flux
to the dust mass as described in Section \ref{subsec:Mdust}.
We adopt $\kappa_\nu$ of silicate. Note that
the dust mass $M_\mathrm{d}$ is proportional to $\kappa_\nu^{-1}$.
For convenience, we list $\kappa_\nu$
normalized to the silicate value (denoted as $\kappa_\mathrm{sil}$)
in Table \ref{tab:kappa},
so that we can multiply the dust mass with $\kappa_\mathrm{sil}^{-1}$
to obtain the dust mass with a different dust species.
In Fig.\ \ref{fig:Mdust_ModelA},
we show the obtained dust mass {corresponding to the
set of ($A_V$, $T_\mathrm{d}$)} by the grey scale.
{We only show the dust mass in the area of
($A_V$, $T_\mathrm{d}$) where the fitting is satisfactory.
The shaded regions show that the reduced $\chi^2$
is larger than 3 or that the mm flux exceeds the ALMA upper limit.}
We obtain a lower dust mass for a higher dust temperature partly
because, as mentioned above (see Fig.\ \ref{fig:Mdust_ModelA}),
the 1.2 mm flux decreases with dust temperature
under a fixed total FIR luminosity (i.e.\ under a fixed $A_V$)
and partly because $B_\nu (T_\mathrm{d})$ is larger for higher
$T_\mathrm{d}$ (see equation \ref{eq:Mdust}).

In Fig.\ \ref{fig:Mdust_ModelA}, we observe that $A_V<0.4$ {mag} is not
allowed. This is because the rest UV--optical SED is too blue with
small extinction. Moreover, the constraint on the dust temperature is
not largely different among various values of $A_V$, because
most of the UV radiation from the youngest stellar population is
absorbed as long as
$A_V\gtrsim 0.5$ {mag}; that is, the FIR luminosity is not sensitive to $A_V$
if $A_V\gtrsim 0.5$ {mag}. Accordingly, the constraint on the dust temperature is only weakly
dependent on $A_V$ as long as $A_V\ga 0.5$ {mag}.
The lower limit for the dust temperature is 70 K at $A_V\sim 0.4$ {mag}
and 80 K at $A_V\sim 1$ {mag}. The dust mass obtained is lower than
$2.1\times 10^6$ M$_{\sun}$. We use this value for the upper limit
of the dust mass in Model A.

\subsection{Himiko: Model B}\label{subsec:ModelB}

We investigate Model B ($\tau_\mathrm{SF}=20$ Myr and $t=200$ Myr)
for Himiko. In this case, the contribution from the stellar population younger
than 10 Myr is negligible, so that the resulting SED is insensitive to
$\eta$. Because ionizing photons are emitted by such a young stellar
population \citep[e.g.][]{Kennicutt:2012aa}, the strong Ly$\alpha$ emission
observed for this galaxy
may not be explained by recent star formation in this model.
However, as discussed in \citet{Ouchi:2009aa}, there are mechanisms
of Ly$\alpha$ emission other than recent star formation such as
ionization by a hidden AGN, cooling of newly accreted gas, outflowing
gas excited by shocks, etc. Therefore, Model~B is still worth investigating,
but we should keep in mind that the strong Ly$\alpha$ emission is
not an indicator of the star formation rate in this case.

The intrinsic stellar SED is almost consistent with the observed
SED in Model B; thus, extinction is not strongly required.
Since the contribution from young stellar population is negligible,
$\eta$ is not important. Thus, we simply
fix $\eta =1$ (so $A_V$ is the extinction for
all stellar populations). We examine which set of
($A_V$, $T_\mathrm{d}$, $\delta$) reproduces the observed SED
with the same procedure as in Model A; that is, we extract
the parameter sets that realizes reduced $\chi^2<3$ and
the flux at 1.2 mm below the 3 $\sigma$ constraint of the ALMA
observation. Below we examine the area of ($A_V$, $T_\mathrm{d}$)
that gives a good fit to the SED under a given $\delta$ ($-0.4$, $-0.7$,
or $-1.1$).

\begin{figure}
 \includegraphics[width=0.98\columnwidth]{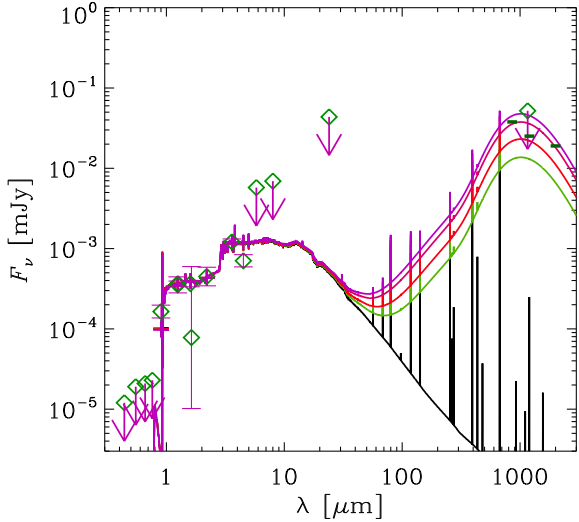}
 \caption{Examples of fitted SEDs for Himiko in Model B with
 $\eta =1$, $T_\mathrm{d}=30$ K, and $\delta =-0.4$. The SEDs show
 the results with $A_V=0$, 0.03, 0.05, 0.08, and 0.1 {mag} for
 the lower to upper lines at 1.2 mm. The data points and the short horizontal
 lines are the same as in
 Fig.\ \ref{fig:Himiko_ModelA}.}
 \label{fig:Himiko_ModelB}
\end{figure}

As examples of satisfactory fitting, we show the SEDs
with $\delta =-0.4$, $T_\mathrm{d}=50$ K, and various $A_V$
in Fig.\ \ref{fig:Himiko_ModelB}. For extinction, $A_V>0.1$ {mag} is
rejected because the UV--optical SED is too red to fit the
observed SED. If we adopt a steeper extinction curve
($\delta =-0.7$ or $-1.1$), the constraint on $A_V$ becomes
more stringent since the UV--optical SED becomes redder with a
smaller $A_V$.

Because $A_V$ is strongly limited by the UV--optical SED in Model B, the FIR
emission is consistent with the ALMA data point even if we adopt a low dust
temperature such as $T_\mathrm{d}=30$ K. In other words, the dust temperature is not
constrained in this model. Therefore, $T_\mathrm{d}$ should be constrained by another
method. H14 suggested that the dust temperature can be estimated by
assuming radiative equilibrium between incident stellar radiation
and dust FIR emission. The obtained equilibrium dust temperature
is $\sim 30$ K; however, H14 argue that this is a lower limit because of
the assumption that dust is distributed over the ALMA beam
(in a radius of 2.2 kpc). In reality, it is expected that the dust is
associated with the stellar distribution (or regions
of recent star formation), which is more compact than
the ALMA beam in Himiko \citep{Ouchi:2013aa}.

In fact, we cannot completely exclude a possibility that the dust
temperature is lower than 30 K. In this case, the dust temperature
is similar to the cosmic microwave background
(CMB) temperature, which is $\sim 20.7$ K at Himiko's redshift ($z=6.6$). 
\citet{Ferrara:2017aa} argue that high-pressure environment in
high-$z$ star-forming galaxies can accommodate
dense regions where the dust is effectively shielded from
the intense stellar radiation \citep[see also][]{Pallottini:2017aa}.
Because the dust is `dark' in this case, even ALMA may not be
able to detect it.

In summary, Model B is reduced to a case of unconstrained dust
temperature. Thus, the best effort we can take is to adopt
the same procedure as in H14 to introduce an additional
constraint on the dust temperature through radiative
equilibrium argument. In this case, the obtained upper limit
for the dust mass is the same as in H14; therefore, we simply
refer to H14 for the dust mass constraint in Model B.

In Fig.\ \ref{fig:Himiko_ModelB}, we also show the expected
ALMA sensitivities (the same as shown in Fig.\ \ref{fig:Himiko_ModelA},
based on 5-hour on-source integration with the full ALMA)
at wavelengths 0.85, 1.2 and 2 mm (frequencies 350, 250, and
150 GHz).
We observe that even with the full ALMA, this object is not detected
unless $A_V>0.05$ {mag}. If $A_V> 0.08$ {mag}, all the three bands can detect
Himiko; in this case, we are able to give a strong constraint on
the dust temperature and
the total FIR luminosity (i.e.\ $A_V$). If the dust temperature is
higher/lower under a fixed $A_V$, the detection at longer wavelengths becomes
more/less difficult.

\subsection{LBG sample}\label{subsec:LBG}

\subsubsection{Fitting to the composite SED}\label{subsubsec:LBG_fitting}

We apply the same fitting procedure as above to the B16 LBG sample.
Although each LBG would put only a much weaker constraint on
the dust production than Himiko, the entire sample may give
a strong constraint after stacking. Thus, we adopt the composite SED
of the B16 LBGs constructed in Section \ref{subsec:data} for the fitting.
Since the age--extinction degeneracy is present,
we apply Models A and B as two `extremes' among the representative cases,
following the fitting to Himiko. As mentioned above, the major difference
between these two
models is the contribution from the
most recent ($\lesssim 10$~Myr) star formation to the intrinsic UV slope.
We regard Model A (continuous SFR) as more probable, since there is no reason
that the LBG sample is
biased to the objects without a recent ($\lesssim 10$ Myr) star formation activity.

Because the ALMA upper limit flux relative to the UV flux is higher than
that of Himiko, the constraint on the parameters is weaker (i.e.\
we find more solutions than in Himiko's case).
The intrinsic (stellar) SED of Model A is bluer than the composite
SED. Thus, we need a significant amount of extinction ($A_V\gtrsim 0.2$ {mag}).
For the steepest extinction curve ($\delta =-1.1$), the upper bound of $A_V$ is about
0.4 {mag}, which is determined by the rest UV SED; if $A_V$ is larger than
0.4 {mag}, the rest UV SED is too red.
If the extinction curve is flatter, we need a higher $A_V$, so that
only high dust temperature is permitted to be consistent with the
strong upper limit at 1.31 mm. Thus, there are two lines of solutions:
(i) one is small $A_V\lesssim 0.4$ {mag} with a steep extinction curve, and (ii) the other
is large $A_V\gtrsim 0.7$ {mag} with a flat extinction curve and a high dust temperature.
Since the composite SED reflects an averaged property for the extinction,
we simply adopt a standard value for $ \eta =0.44$, appropriate for
nearby starburst galaxies \citep{Calzetti:2000aa}.
We also examine the case of $\eta =0$ later.

If we use reduced $\chi^2<3$ as a criterion of good fit following
the case of Himiko, it also allows systematically
redder SEDs that are not consistent with the flat wavelength
dependence in the composite rest UV SED. We also found that
the SED at $\lambda > 3~\micron$ does not constrain
the parameters because the accepted flux ranges are wide.
Thus, we only choose solutions that are within the shaded
region at $\lambda <3~\micron$ and below the 3 $\sigma$
upper limit at 1.3 mm in Fig.\ \ref{fig:LBGSED}.

\begin{figure}
 \includegraphics[width=1\columnwidth]{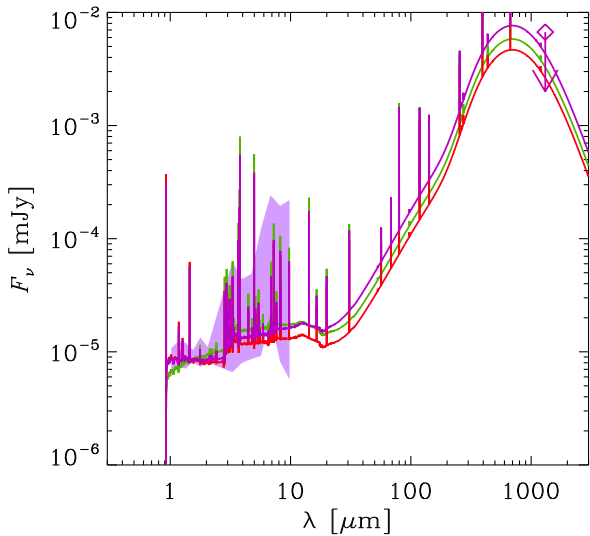}
 \caption{Examples of the fitting to the composite SED of the B16 LBG sample.
 Model A ($\tau_\mathrm{SF}=2000$ Myr and $t=400$ Myr) is adopted with $\eta =0.44$.
 The SEDs show the results with $(A_V\,\mathrm{[mag]},\,\delta )=(0.7,\,-0.4),\, (0.4,\,-1.1)$
 and $(1,\, -0.4)$ for the lower to upper lines at mm wavelengths.
 The point with an arrow at 1.3 mm is a 3 $\sigma$ upper limit obtained by stacking,
 while the shaded region at $\lesssim 10~\micron$ shows the {probable}
 area of the composite SED which is also shown in Fig.\ \ref{fig:LBGSED}.
 Note that we only used the data at $\lambda <3~\micron$ and
 the mm upper limit for the fitting {(see the text)}.}
 \label{fig:fitLBG}
\end{figure}

In Fig.\ \ref{fig:fitLBG}, we show case (i) with $\delta =-1.1$.
We show $A_V=0.4$ {mag} with $T_\mathrm{d}=50$ K as an example of
satisfactory fit.
In the same figure, we also show
case (ii) with $\delta =-0.4$, adopting $A_V=0.7$ and 1 {mag}
with the same dust temperature as above ($T_\mathrm{d}=50$ K).
In case (ii), the rest UV SED is consistent with the composite SED
even for $A_V=1$; however, such a large $A_V$ predicts large
FIR luminosity, so that high dust temperature is required to be consistent
with the ALMA upper limit.
The rest UV SED ($\lambda <3~\micron$) is still as blue as
seen in the composite SED because the extinction curve is flat.
Since the total dust emission luminosity is higher than in case (i),
the dust temperature should be $\gtrsim 50$ K in case (ii) if
$A_V$ is as large as 1 {mag}.
Therefore, the constraint on the dust temperature is tight
for a flat extinction curve.

We also confirmed that $\eta =0$ gives the satisfactory fit to
the UV SED as long as $A_V>0.2$ {mag} (with $\delta =-1.1$;
see also Section \ref{subsubsec:Mdust_LBG}) in Model A.
Therefore, the above assumption of
$\eta =0.44$ is not essential, but extinguishing the bluest stellar
population with age $<10$ Myr is essential to reproduce the UV
colour of the composite SED. It is not probable that the major part of
the LBGs has stopped star formation in the last 10 Myr; thus, the
dust extinction is the only probable way of systematically eliminating
the contribution from the youngest ($<10$ Myr) population.

Nevertheless, we still investigate Model B  ($\tau = 20$ Myr and $t= 200$ Myr),
in which the population with
age $<$10 Myr has negligible contribution to the {intrinsic} SED, because it is
difficult to completely reject this model.
For Model B, since the intrinsic stellar SED is already consistent with the
observed SED, only $A_V<0.1$ {mag} is permitted. 
With such a low extinction value, the FIR emission is well below the
ALMA upper limit. However, as mentioned above, we regard Model B
as improbable for the stacked data. 
Thus, we focus on Model A for the B16 sample.

\subsubsection{Constraint on the dust mass}\label{subsubsec:Mdust_LBG}

\begin{figure*}
 \includegraphics[width=1\columnwidth]{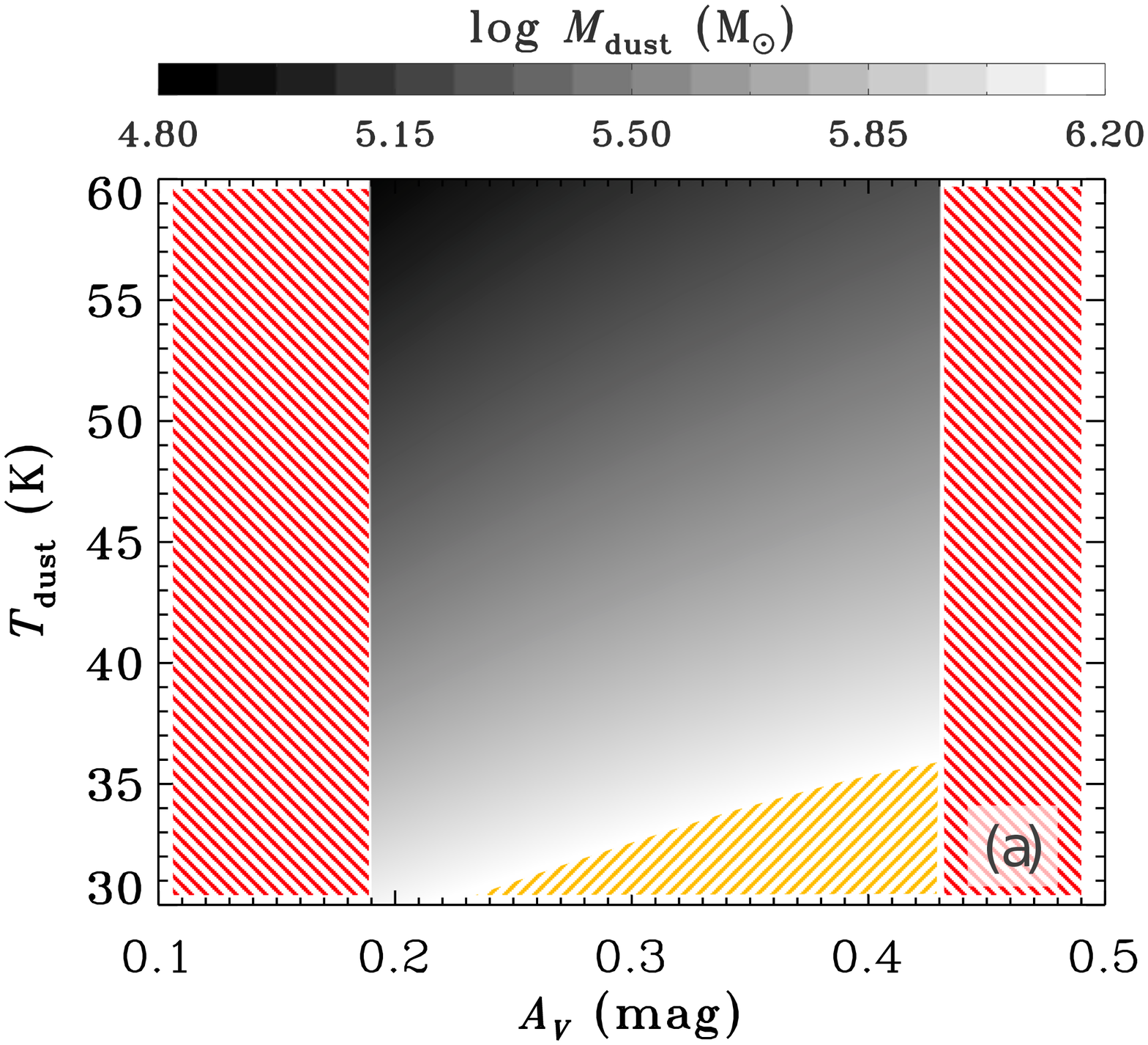}
 \includegraphics[width=0.9\columnwidth]{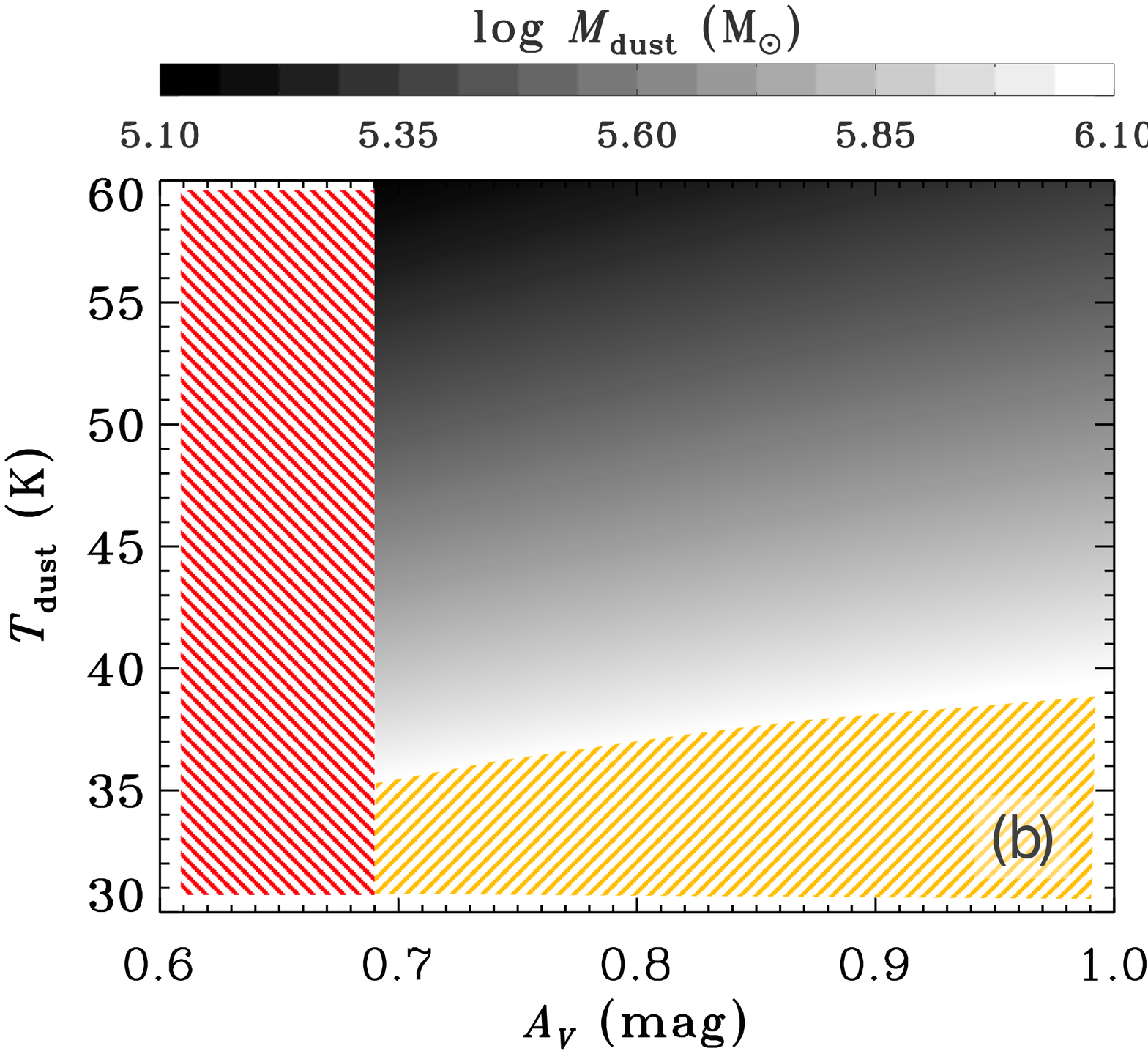}
 \caption{Same as Fig.\ \ref{fig:Mdust_ModelA} but for the dust mass obtained
 for the composite LBG SED.  A steep extinction curve with $\delta =-1.1$
 and a flat extinction curve with $\delta =-0.4$ are adopted for panels (a) and
 (b), respectively. For both panels,
 $\eta =0.44$ is adopted.}
 \label{fig:Mdust_LBG_A}
\end{figure*}

We estimate the dust mass for the allowed
parameter ranges as already done for Himiko in Fig.\ \ref{fig:Mdust_ModelA}
(Section \ref{subsec:ModelA}). In Fig.\ \ref{fig:Mdust_LBG_A},
we show the dust mass on the $(A_V,\, T_\mathrm{d})$ plane
for $\delta =-1.1$ and $-0.4$ as representatives of steep (SMC-like)
and flat extinction curves, respectively (Section \ref{subsec:cigale}).
We adopt the same dust species ($\kappa_\nu$) as adopted for
Himiko in Section \ref{subsec:ModelA}.

We observe that $0.19\,\mathrm{mag}<A_V<0.43\,\mathrm{mag}$ is allowed for the steep extinction
curve while $0.69\,\mathrm{mag}<A_V$ is allowed for the flat extinction curve.
The flat extinction curve requires a higher extinction to make the
rest UV--optical SED as red as observed. Because the reddening is small,
extinction as large as $A_V>1$ {mag} is allowed in this case.
Therefore, if a major part of LBGs have as flat an extinction curve as
observed in high-$z$ quasars, the extinction should be
large, so that the correction for stellar mass and star formation rate for
the extinction is as large as a factor of $\gtrsim 3$ even at $z>5$.
If such a large fraction of star formation activity is enshrouded by dust,
there should be a discrepancy between the time integration of the star formation
rate traced in the UV and the observed stellar mass. However,
there is no strong evidence of such discrepancy at $z>5$
\citep{Madau:2014aa}.
Therefore, we judge that the steep extinction curve with
small $A_V\sim 0.2$--0.4 {mag} is more probable, although we should
keep in mind that a flat extinction curve with large $A_V$ could also give
a consistent SED with observed LBGs at $z\gtrsim 5$.
The dust temperature should be higher than 35 K for the flat extinction
curve while dust temperatures as low as 30 K are allowed for the
steep extinction curve. This is because large dust extinction
for the flat extinction curve predicts a high dust emission luminosity,
in which case the SED peak should be located at
shorter wavelengths than 1.31 mm for the consistency with the ALMA upper limit.
Overall, the constraint on the dust temperature is weaker compared with
the case of Himiko, since the upper limit at the ALMA band relative to the
UV continuum level is higher for the LBGs than for Himiko.

\section{Constraint on dust production}\label{sec:SNdust}

H14 (see their section 3) proposed a method of constraining the dust
production per SN. They basically divide the total dust mass by
the number of SNe. This obtained dust mass per SN is an upper limit
in the following two senses: (i) the estimate is based on an upper limit of
the dust emission flux;
and (ii) it is based on the assumption that all the dust is produced by
SNe (i.e.\ it neglects other formation paths of dust). In this section,
we briefly review their method and apply it to the dust masses obtained
above for Himiko and LBGs.

\subsection{Method of constraining SN dust production}\label{subsec:SNdust}

We constrain the dust mass formed in a single SN by using
the upper limit dust masses obtained above. If we assume that
all the dust originates from dust condensation in SN ejecta,
we can estimate the dust mass ejected from a single SN, $m_\mathrm{d,SN}$ as
\begin{align}
m_\mathrm{d,SN}=\frac{M_\mathrm{d}}{(1-f_\mathrm{dest})N_\mathrm{SN}},
\label{eq:mdSN}
\end{align}
where $f_\mathrm{dest}$ is the fraction of dust destroyed by SN shocks in the ISM, and $N_\mathrm{SN}$ is the total number of SNe
\citep[see also][]{Michalowski:2015aa}. We neglect the effect of dust recycling
in star formation, since the assumption that SNe are the dominant
source over grain growth indicates an early stage of chemical evolution
\citep{Dwek:1998aa,Inoue:2011aa,Zhukovska:2008aa}.
To obtain $m_\mathrm{d,SN}$, we need to estimate
$N_\mathrm{SN}$ and $f_\mathrm{dest}$.

The total number of SNe at age $t$, $N_\mathrm{SN}(t)$, is estimated by
\begin{align}
N_\mathrm{SN}(t) &= \int_0^t
\int_{8~\mathrm{M}_\odot}^{40~\mathrm{M}_{\sun}}
\,\psi (t'-\tau_m)\phi (m)\,\mathrm{d}m\,\mathrm{d}t'\nonumber\\
&\simeq \int_0^t\psi (t')\,\mathrm{d}t'
\int_{8~\mathrm{M}_\odot}^{40~\mathrm{M}_{\sun}}\phi (m)\mathrm{d}m,
\label{eq:N_SN}
\end{align}
where $\psi (t)$ is the star formation rate at $t$,
$\tau_m$ is the lifetime of a star with mass $m$
(mass at the zero age main sequence),
$\phi (m)$ is the IMF, and stars in the mass range of
8--40 $\mathrm{M}_{\sun}$ are assumed to evolve into
SNe \citep{Heger:2003aa}.
We assume that the lifetimes
of SN progenitors are much shorter than $t$ in order to
simplify the first line of equation (\ref{eq:N_SN}) to the second. 
For the consistency with the above SED fitting, we adopt
the Chabrier IMF. The IMF is normalized
so that the integral of $m\phi (m)$
for the entire mass range is 1.

The integration for the IMF in equation (\ref{eq:N_SN}) indicates
the number of SN progenitors per stellar mass, and is denoted
as $\mathcal{F}_\mathrm{SN}$:
\begin{align}
\mathcal{F}_\mathrm{SN}\equiv
\int_{8~\mathrm{M}_\odot}^{40~\mathrm{M}_{\sun}}\phi (m)\mathrm{d}m.
\end{align}
We estimate that $\mathcal{F}_\mathrm{SN}=9.9\times 10^{-3}$~M$_{\sun}^{-1}$
for the IMF adopted.
The other factor in equation (\ref{eq:N_SN}) is the integrated star formation
rate and is denoted as $\mathcal{M}_*$:
\begin{align}
\mathcal{M}_*\equiv\int_0^t\psi (t')\,\mathrm{d}t'.
\end{align}
We are able to derive $\mathcal{M}_*$ from the SED fitting as an output
quantity. Using the above two quantities, equation (\ref{eq:N_SN}) is reduced to
\begin{align}
N_\mathrm{SN}=\mathcal{F}_\mathrm{SN}\mathcal{M}_*.\label{eq:N_SN_reduced}
\end{align}

H14 used a dust evolution model to estimate
$f_\mathrm{dest}$. The dust destruction is most prominently
seen at the metallicity level where the dust growth by accretion
starts to dominate the dust abundance. Thus, if we use the
destroyed fraction at this metallicity level, we are able to obtain
the most conservative (i.e.\ largest) value for $f_\mathrm{dest}$.
Following their estimate, we adopt $f_\mathrm{dest}=0.5$.

\subsection{Constraint on SN dust for Himiko}

In Section \ref{subsec:ModelA}, we have shown that, if we adopt
Model A (a continuous SFR) for Himiko, only high dust temperatures
($T_\mathrm{d}>70$~K) are allowed to make the dust emission SED
consistent with the ALMA upper limit. Because of the high dust
temperature as well as the tight ALMA upper limit, we obtain
a stringent upper limit of $2.1\times 10^6$ M$_{\sun}$ for the total dust mass
(Section \ref{subsec:ModelA}). H14 obtained an upper limit of
$2.0\times 10^7$ M$_{\sun}$ for the same dust material
(silicate). The difference arises from their different method of
estimating the dust temperature: H14 derived the dust mass
based on the radiative equilibrium argument and obtained
$T_\mathrm{d}\sim 30$--40 K. As already discussed in
Section \ref{subsec:ModelB}, this dust
temperature may be an underestimate because their estimate of
the dust heating rate is based on the assumption that the
dust is extended over the ALMA beam.
{Thus, the higher dust temperatures than obtained by H14
are not inconsistent with the current observational knowledge
for Himiko.}

Now we apply the method described in Section \ref{subsec:SNdust}
to obtain a constraint on the dust mass per SN. We estimate the integrated
star formation rate as
$\mathcal{M}_*=10^{9.8}$--$10^{9.9}$ M$_{\sun}$ for the SEDs with
satisfactory fitting.
Thus, using equation (\ref{eq:N_SN_reduced}), we obtain the total number of
SNe as $N_\mathrm{SN}=(6.2\mbox{--}7.9)\times 10^7$.
Using equation (\ref{eq:mdSN}) and recalling that $f_\mathrm{dest}=0.5$,
we finally obtain $m_\mathrm{d,SN}<0.053$--0.067~M$_{\sun}$
based on the upper limit of $M_\mathrm{d}$ ($2.1\times 10^6$ M$_{\sun}$).
We take the larger value as a conservative limit
(0.067 M$_{\sun}$). The dust masses obtained for other dust species are
listed in Table \ref{tab:mdust_Himiko}.

\begin{table*}
\centering
\begin{minipage}{105mm}
\caption{Upper limits for the total dust mass ($M_\mathrm{d}$) and dust mass
produced per SN ($m_\mathrm{d,SN}$).}
\label{tab:mdust_Himiko}
    \begin{tabular}{lcccccc}
     \hline
     & \multicolumn{2}{c}{Himiko Model A} & \multicolumn{2}{c}{Himiko Model B$^\mathrm{a}$}
     & \multicolumn{2}{c}{LBG Model A}\\
     Species$^\mathrm{b}$ & $M_\mathrm{d}$ & $m_\mathrm{d,SN}$ &
     $M_\mathrm{d}$ & $m_\mathrm{d,SN}$ &
     $M_\mathrm{d}$ & $m_\mathrm{d,SN}$\\
      & ($10^6$ M$_{\sun}$) & (M$_{\sun}$) & ($10^6$ M$_{\sun}$) & (M$_{\sun}$)
      & ($10^6$ M$_{\sun}$) & (M$_{\sun}$) \\
     \hline
     Graphite & 1.3 & 0.042 & 14 & 0.18 & 1.0 & 1.4\\
     Silicate & 2.1 & 0.067 & 20 & 0.25 & 1.6 & 2.3\\
     SN$_\mathrm{con}$ & 5.0 & 0.16 & 27 & 0.34 & 3.8 & 5.5\\
     SN$_\mathrm{dest}$ & 3.2 & 0.10 & 22 & 0.28 & 2.4 & 3.4\\
     AC  & 0.97 & 0.031 & 8.9 & 0.11 & 0.74 & 1.0\\
     \hline
    \end{tabular}
    
    \medskip

$^\mathrm{a}$Because the dust temperature is not constrained in this model,
the dust mass is not well determined. Thus, we put the dust mass
constraint obtained using the radiative-equilibrium dust temperature
derived by H14 (the total dust mass is the same as {in} their paper
while the dust mass per SN is modified because we adopted a different IMF
{and stellar mass}).\\
$^\mathrm{b}$See Table \ref{tab:kappa} and the text for the dust species.
\end{minipage}
\end{table*}

For Model B, because the dust temperature is not constrained by
our method, we simply adopt the dust mass obtained by H14 as mentioned
in Section \ref{subsec:ModelB}.
{
The integrated stellar mass is $\mathcal{M}_*=10^{10.2}$--$10^{10.3}$~M$_{\sun}$
in Model B of Himiko.
This leads to the number of SNe as $N_\mathrm{SN}=(1.6$--$1.8)\times 10^8$.
Using equation (\ref{eq:mdSN}) together with $f_\mathrm{dest}=0.5$,
we obtain $m_\mathrm{d,SN}<0.22\mbox{--}0.25$ M$_{\sun}$ for silicate
based on the upper limit of $M_\mathrm{d}$ ($2.0\times 10^7$ M$_{\sun}$).
We take the larger value as a conservative limit
(0.25 M$_{\sun}$).}
The upper limit obtained for each dust species is listed
in Table \ref{tab:mdust_Himiko}.

\subsection{Constraint on SN dust for LBGs}

We apply the same method as above to constrain the dust mass produced
per SN for the LBG sample in B16.
Because Model B does not put a meaningful constraint on
$M_\mathrm{d}$ (Section \ref{subsubsec:LBG_fitting}),
we concentrate on Model A. In Section \ref{subsec:LBG},
we used the composite SED to constrain the total dust mass, obtaining
an upper limit of $\sim 1.6\times 10^6$ M$_{\sun}$ (we adopt
$\delta =-1.1$, since this gives a more conservative upper limit than
$\delta =-0.4$) for silicate. In Table \ref{tab:mdust_Himiko},
we list the corresponding upper limits for the different species.
For the composite SED, we obtain
the integrated star formation rate as $\mathcal{M}_*=1.4\times 10^8$~M$_{\sun}$.
Thus, from equation~(\ref{eq:N_SN_reduced}), we obtain $N_\mathrm{SN}=1.4\times 10^6$.
Using equation (\ref{eq:mdSN}) along with $f_\mathrm{dest}=0.5$,
we finally obtain an upper limit of $m_\mathrm{d,SN}<2.3$ M$_{\sun}$.
The dust mass obtained for other dust species are
listed in Table \ref{tab:mdust_Himiko}.

The upper limit of $m_\mathrm{d,SN}$ obtained for the LBGs is too large
to put a useful constraint on the SN dust production theory.
This weaker constraint than in the case of Himiko arises
from the much smaller number of SNe ($N_\mathrm{SN}$).
Therefore, in order to obtain a strong constraint on $m_\mathrm{d,SN}$,
it is desirable to observe a system in which a large number of SNe
have occurred. Because the number
of SNe is proportional to the total stellar mass (with a fixed
IMF), observations of objects with high (stellar) UV luminosity
give strong constraint on $m_\mathrm{d,SN}$.

\section{Discussion}\label{sec:discussion}

We have shown that we are able to put a strong constraint on
the dust mass in Himiko. There are two ways
of reproducing the observed rest-UV colour:
one is to apply a blue SED with significant dust extinction for the youngest
($\lesssim 10$ Myr) stellar population, and the other is to assume a stellar
population which lacks the youngest population.
The first and second cases correspond to Models A and B, respectively.
The derived dust mass depends on which model to adopt.
Although we are not able to draw a definite conclusion regarding
which model is correct for Himiko or LBGs, we here mainly discuss
how we will be able to distinguish between
the two models.

\subsection{How to distinguish between Models A and B}\label{subsec:distinguish}

The largest difference between Models A and B is the magnitude of
dust extinction. Model A, which requires more extinction than Model B,
tends to predict higher total dust luminosities; thus, for the consistency
with the ALMA upper limits, higher dust temperatures are required
in Model A than in Model B. For Himiko, we obtained dust temperatures
higher than 70 K {in Model A}. With such a high temperature, the SED peak is located
at a shorter wavelength than the ALMA 1.2 mm band. Therefore, if we
observe Himiko at a shorter wavelength such as at 850 $\micron$, we
could see if the high dust temperature is really the solution as we already
discussed in Section \ref{sec:result}.

For the stacked SED of the B16 LBG sample, dust temperatures
as low as $\sim 40$ K are still allowed (Fig.\ \ref{fig:Mdust_LBG_A}).
The weak constraint is due to the faint UV flux.
Because the UV flux of {an} LBG is
on average 30--50 times lower than that of Himiko, we need
$\sim 30^2$--$50^2$ LBGs to obtain as strong constraint
as we obtained for Himiko. In order words, the observation of
an object whose UV flux is as bright as Himiko is 1000--3000 times
more powerful in terms of the integration time than that of a normal LBG
in constraining the SN dust production.

\subsection{High dust temperature}

In the above, we suggested a high dust temperature for Himiko.
Indeed, some studies have suggested that the dust temperatures
in high-$z$ star-forming galaxies are high.
\citet{Ouchi:1999aa} showed that dust temperatures should be
higher than 40 K for $z\sim 3$ LBGs if the lack of detection by SCUBA
is taken into account. B16 derived similarly high dust temperatures
for LBGs at $z\gtrsim 5$ based on the
deficit of ALMA detection.

There are also some theoretical studies that suggested high temperatures
in high-$z$ galaxies.
\citet{Ferrara:2017aa} estimated that the dust temperature in LBGs could
be as high as $\gtrsim 50$ K if the spatial distribution of dust is as compact
as the stellar distribution.
\citet{Narayanan:2017aa}, using their cosmological hydrodynamic simulation
and radiation transfer calculation, showed that the dust temperatures in
high-$z$ dusty star-forming galaxies are as high as 50--70 K.
These studies give a physical reason for the high
dust temperatures at high $z$.

There are some LBGs at $z>5$ whose dust continuum was detected by ALMA
\citep{Capak:2015aa,Watson:2015aa,Laporte:2017aa}.
Although we need careful one-by-one analysis for those galaxies,
our preliminary results (Burgarella et al., in preparation) indicate that
they tend to have high dust temperatures
{\citep[see][for a very recent result]{Faisst:2017aa}}. Future more sensitive observations
by the full ALMA will also serve to detect more high-$z$ LBGs
at multiple submm--mm wavelengths, enabling us to determine the
dust temperatures. This leads to solving the degeneracy between age
and dust extinction.

\subsection{Constraint on the SN dust production}

In Section \ref{sec:SNdust}, we constrained the dust mass produced
per SN. We obtained a strong upper limit as
$m_\mathrm{d,SN}\lesssim 0.1$~M$_{\sun}$ for Himiko if we adopt Model A.
If we recall the discussion in Section \ref{sec:SNdust}, the constraint
on $m_\mathrm{d,SN}$ was obtained from the dust mass divided by
the number of SNe. The number of SNe was derived from the time
integration of SFR. Although the age and the SFR are uncertain
in the SED fitting, the estimate of integrated SFR is robust
{since it is determined by the observed stellar continuum level}.
Therefore, the smallness of
$m_\mathrm{d,SN}$ is a robust conclusion derived from the
ALMA upper limit and the rest-UV flux.

The small $m_\mathrm{d,SN}$ indicates either that
reverse shock destruction in SNe is efficient because of high
ambient medium density ($\gtrsim 10$ cm$^{-3}$)
\citep{Nozawa:2007aa,Bianchi:2007aa}, or that
the dust is lost for some reason such as galactic winds, etc.
The possibility of high ambient density may be supported by
\citet{Pallottini:2017aa}, who showed based on their hydrodynamic simulation
that the central gas disc of a high-$z$ galaxy has a density higher than 25 cm$^{-3}$.
In such a dense environment, reverse shock destruction could be efficient.
For the latter possibility, \citet{Hou:2017aa}, based on the numerical simulation
developed by \citet{Aoyama:2017aa}, showed that dust can be transported
into the circum-galactic space by SN feedback.
\citet{Bekki:2015aa} suggested that the dust loss by stellar feedback
could be important in explaining the extinction curve in the SMC.

The above scenario of low $m_\mathrm{d,SN}$ implies that the
dust enrichment by SNe is not efficient at high $z$. Probably other mechanisms
such as dust production by AGB stars \citep{Valiante:2009aa} and
dust growth in the dense ISM \citep{Mancini:2015aa,Popping:2016aa,Wang:2017aa}
are necessary to
produce an appreciable amount of dust at high $z$
{\citep[see also][]{Ferrara:2016aa,Zhukovska:2016aa}}.
We note that, even if we consider other processes of dust formation,
the values of $m_\mathrm{d,SN}$ obtained above are still an upper limit because we
assumed that all dust is produced by SNe.

\section{Conclusion}\label{sec:conclusion}

We investigate the possibility of constraining the dust mass in
high-redshift ($z\gtrsim 6$) galaxies by applying SED fitting (\textsc{cigale})
to rest UV--optical photometric data and the ALMA upper limits.
For SED fitting, there is a well known degeneracy between dust extinction
and stellar age.
Moreover, the Ly$\alpha$ emission line is not necessarily associated
with star formation activity, which means that the bright Ly$\alpha$ emission
cannot completely exclude a possibility of old stellar age.
Therefore, we focus on two extremes for the star formation history:
one is continuous star formation which includes very young
($<10$ Myr) population (Model A), and the other is instantaneous star formation
in which the star formation rate has declined (with negligible young stellar
population with ages $<10$ Myr; Model B). These models are applied
to Himiko and the B16 LBG sample.

For Himiko, Model A predicts significant dust extinction to explain
the observed rest-UV SED.
The predicted 1.2 mm flux is consistent with the strong ALMA upper limit
only if the dust temperature is higher than 70 K. Because of
the high dust temperature (i.e.\ high emission efficiency), we obtain
a strong upper limit for the dust mass $\sim 2\times 10^6$ M$_{\sun}$.
Based on this value, the dust mass produced per SN is estimated as
$\lesssim 0.1$ M$_{\sun}$. This low value indicates that dust once condensed
is destroyed in the shocked region associated with the SN, or that dust is
lost out of the main body of the galaxy. If this is true for other galaxies
at high $z$, SNe may not be the main source of dust there,
{and} we need to consider other processes for dust enrichment
such as dust growth in the dense ISM. In contrast, Model B allows
an order of magnitude {larger} dust mass $\sim 2\times 10^7$ M$_{\sun}$,
which is converted to the dust mass produced by a SN as
$\sim 0.3$ M$_{\sun}$. We could distinguish between Models A and B
if we observe Himiko at a shorter wavelength than 1.2 mm by the full ALMA.
The high dust temperatures in Model A predict that Himiko
can be detected at 0.85 mm.

For the LBG sample in B16, we make a composite SED to put
a strong constraint on the ALMA mm flux. The composite SED indicates
that the dust mass is $\sim 2\times 10^6$ M$_{\sun}$ or less in
a typical LBG at $z>5$, but this only puts a weak upper limit for
the dust mass produced per SN as $\lesssim 2$ M$_{\sun}$.
We estimate that, in order to obtain an upper limit comparable to Himiko
for SN dust production,
we need to observe 1000--3000 LBGs. This clarifies the importance
of observing UV-bright objects (like Himiko) to constrain
the dust production by SNe.

\section*{Acknowledgements}

We are grateful to the anonymous referee for useful comments.
HH thanks the staff at LAM for their hospitality and financial and technical
support during my stay.
HH is supported by the Ministry of Science and Technology
grant MOST 105-2112-M-001-027-MY3.



\bibliographystyle{mnras}
\bibliography{hirashita}




\bsp	
\label{lastpage}
\end{document}